# Density Functional Theory Study of Light Metal (Li/Na/Ca) Functionalized Borophosphene for Reversible Hydrogen Storage

Sandip Haldar[a,*]

[a]School of Mechanical Sciences, Indian Institute of Technology Goa, Farmagudi, Goa, India 403401

Email: sandip@iitgoa.ac.in

**Abstract**

Borophosphene is investigated for hydrogen storage by density functional theory calculations through Li, Na and Ca decoration. Decoration enhances the binding energy from -0.047 eV/$H_2$ to -0.20 – -0.42 eV/$H_2$. PDOS and Bader charge analysis elucidate the role of adatom decoration in charge transfer and better binding. Up to 10, 12 and 20 $H_2$ molecules can be adsorbed over a single Li, Na and Ca adatom, respectively, in a supercell of 32 atoms. Desorption temperature is calculated from the binding energies. A complete discharge of the stored molecules from decorated borophosphene can be realized in temperature range of 125 – 531 K. Further, decoration at multiple sites of the substrate is performed to evaluate the theoretical gravimetric density. With Li, Na, and Ca overloading, gravimetric densities of 6.22%, 5.34%, and 6.08% are obtained. NEB results show that inter-site energy barriers of the adatoms are larger than their thermal energy by an order.

## 1. Introduction

Hydrogen energy is one of the most available clean energy solutions due to its pollution free nature and high energy density per unit weight and brings the potential to alleviate the carbon footprint from the fossil fuels [1]. Conventional approaches, e.g. pressurized tank and liquid hydrogen fuel come with safety concerns, higher cost along with inadequate energy density [2, 3]. Molecular hydrogen storage over nanomaterials is one of the sought after solutions for hydrogen energy. The US Department of Energy (DOE) target is set at 5.5 – 9.5 % gravimetric density by 2025 and a binding energy between physiosorption and chemisorption [4–6]. Nazir et al. [7] reviewed the challenges and state of the art in $H_2$ storage and outlook toward $H_2$ based green energy. 2D materials have propelled a wide attention in design and synthesis from several novel elements since the grand arrival of graphene. Due to their high specific area, several surface dominated applications have been considered to be beneficial including hydrogen storage. A wide variety of monoelemental 2D materials have been thoroughly scrutinized for their application in molecular hydrogen storage, for example, carbon allotropes [3, 8–12], allotropes of phosphorous [13–18], allotropes of boron [19–23], silicene and germanene [2, 24, 25] etc. Apart from monoelemental 2D materials, different dielemental 2D materials have been considered for hydrogen storage, for example, Boron Nitride [26–28], Boron sulfide [29], Zinc oxide [30], magnesium hydride [31], Beryllium polynitrides [32], Boron/Carbon nitride [5, 33] etc.

Generally, pristine materials exercise poor interaction with the $H_2$ molecules resulting in weak binding energy that is unsuitable for reversible hydrogen storage [4, 33]. For example, pristine graphene, phosphorene, borophene show binding energy of 0.04 - 0.10 eV/$H_2$ in storing molecular $H_2$ due to weak interaction [12, 15, 21, 34, 35]. One of the popularly adopted strategies to enhance the interaction, and thereby, hydrogen storage performance, have been adatom decoration, defect engineering, or both. For decoration (also referred as surface functionalization), alkali metals or transition metals are widely chosen to enhance the interaction through the contribution of the adatoms [20, 21, 34, 36–39]. Metal decorations with low electronegativities become strongly polarized after being adsorbed over the substrate and as a result, attract $H_2$ molecules [3]. Functionalization of the 2D materials (e.g. decoration) enhanced the substrate-$H_2$ interaction by the charge transfer and was adopted as a promising approach for im- proving the binding with $H_2$ molecule. This, as a result, improved the storage capacity or gravimetric density of $H_2$ storage onto the 2D substrates. By decoration, the binding energy of $H_2$ molecules are improved multi-fold and enhance the gravimetric density. The alkali metals offer the binding energy for physiosorption through charge polarization and minimize the cluster formation. On the other hand, tran- sition metals offer stronger binding energy by Kubas type interaction where metal-$d$ and H$_2$-$s$ participate in hybridization, however, they tend to cluster [29, 40, 41]. Further, while transition metals can adsorb more $H_2$ molecules, they compromise the gravimetric density due to their higher atomic mass.

The promise and high performance of the 2D materials have led to further efforts in design and search of novel materials as well as computational screening of their performance while fabrication of them is awaited. Since its prediction, borophosphene has recently garnered a wide interest for different applications such as energy applications. A anisotropic Dirac material with graphene like hexagonal structure constituting of Phosphorous (P) and Boron (B) atoms, referred as borophosphene, was proposed by Zhang et al. [42] and its stability was established. A structure in Pmmm plane group, with B-P-P-B sequence, the unit cell has two lattice constants as 3.22 Å in zigzag direction and 5.57 Å in the armchair direction [42, 43]. Experimental realization is foreseen from the favorable 4.82 eV/atom cohesive energy and 12 meV/Å$^2$ energy for exfoliation [42, 44]. Borophosphene has been evaluated for lithium-, non-lithium-ion batteries, lithium-sulfur, sodium-sulfur batteries [44–49].

This work reports hydrogen storage performance of borophosphene as evaluated using Density Functional Theory (DFT). Light metal (Li, Na and Ca) decoration has been investigated to enhance the hydrogen storage performance of borophosphene by calculating the binding energies. The results have been compared with the monoelemental counterparts, i.e. borophene and phosphorene. Finally, theoretical gravimetric density has been calculated from the results. Further, competition between decoration and clustering of the adatoms over the borophosphene was checked from diffusion barrier energy of the adatoms from the neighboring adatom.

2. **Material and Computational detail**

The substrate is taken as a supercell of 32 atoms from 4 × 2 unit cells and was subjected to energy minimization to determine the ground state structure. $H_2$ was added to the substrate and binding energy was obtained from the energy minimization of the system. For functionalization, Li, Na and Ca decoration over the relaxed pristine substrate were first stabilized and then, computations for hydrogen storage over the decorated substrate were performed.

Quantum Espresso, available under GNU license, was used for the DFT calculations [50, 51]. PAW pseudopotentials were used to model the elements along with Perdew-Burke-Ernzerhof (PBE) exchange correlation functional [52, 53]. The pseudopotentials treat B: $2s^22p^1$, P: $3s^23p^3$, Li: $1s^22s^1$, Na: $2s^22p^63s^1$, Ca: $3s^23p^64s^2$, H: $1s^1$ as valence electrons. The van der Waals forces were corrected through DFT-D2 framework [54]. The wave functions are truncated at a cut-off of 60 Ry and 480 Ry is used for charge density cut-off $1e^{-5}$ Ry is used as convergence threshold for the total energy of the system for SCF calculations. Brillouin zone integration was performed using Monkhorst-Pack grid with $9 \times 7 \times 1$ k-points [45, 55]. Degauss value of 0.02 in Methfessel-Paxton smearing was used in the simulations [56]. In all cases, at least 15 Å vacuum space was added above the substrate to eliminate interlayer long range interactions that may arise from the periodic image.

Hydrogen storage of borophosphene is evaluated by calculating the binding energy given by,

$$E_b^{H_2} = (E_{BP+nH_2} - E_{BP} - E_{nH_2})/i, \tag{1}$$

$$E_b^{H_2} = (E_{BP+LM+nH_2} - E_{BP+LM} - E_{nH_2})/i, \tag{2}$$

for pristine and decorated borophosphene, respectively. The total energy $E_*$ of $*$ material system is obtained from the DFT results, and $n$ is number of $H_2$ molecules in the system. In order for borophosphene to adsorb the $H_2$ molecules this energy has to be negative. A higher magnitude of $E_b$ implies that the substrate binds strongly. To enquire the interaction among the elements, projected density of states (PDOS) are presented and Bader charge analysis for quantitative analysis. Visualization figures are prepared using XCrySDen package [57].

## 3. Results and discussions

### 3.1. Borophosphene

For completeness and validation, pristine borophosphene is first studied to compare with results reported in the literature. The optimized borophosphene is shown in Figure 1(a) along with band structure in Figure 1(b). The conventional unit cell with four atoms is also marked in the figure. Two relevant locations, that will be referred later, are denoted by $H_B$ and $H_P$, in the ring of four B atoms, and four

P atoms, respectively. The substrate of 32 atoms 4 × 2 cells was adopted from the previous reports in literature [44, 45]. The lattice parameters from the relaxed structure were measured as $a$ = 3.22 Å, and $b$ = 5.56 Å. The lengths of the B-B, B-P and P-P bonds are stable at 1.66 Å, 1.84 Å, and 2.10 Å, respectively. These parameters are very close to those reported in literature [42]. With a Dirac cone between Γ and X points, the band structure is also similar to those reported earlier [42].

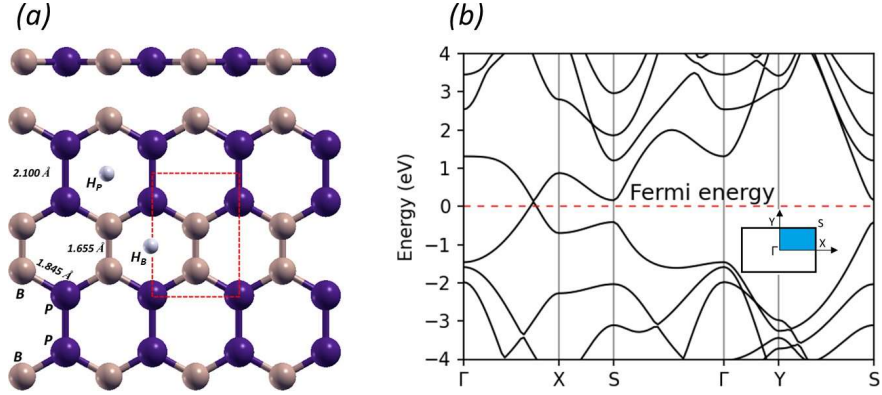

Figure 1: Borophosphene: (a) relaxed monolayer of 4×2 unit cells, and (b) band structure.

## 3.2. $H_2$ storage in borophosphene

Borophosphene and the $H_2$ molecule together was stabilized from a number of unique locations of $H_2$ as indicated in Figure 2(a). Binding energies (Eq. 1) for those initial locations were obtained between -0.029 to -0.047 eV/$H_2$. The best binding configuration ($E_b$ = −0.047 eV) is shown in Figure 2(b) resulting in $H - H$ bond of 0.751 Å and positioned at a height of 2.92 Å from the substrate. Borophosphene exhibits weak binding with $H_2$ in comparison with graphene ($E_b$ = −0.10 eV) [12, 35], however, closer to borophene ($E_b$ = −45 meV) [21], and phosphorene ($E_b$ = −70 meV) [34]. Binding energies of $H_2$ molecule over different 2D materials are compared in Table 1 from literature data.

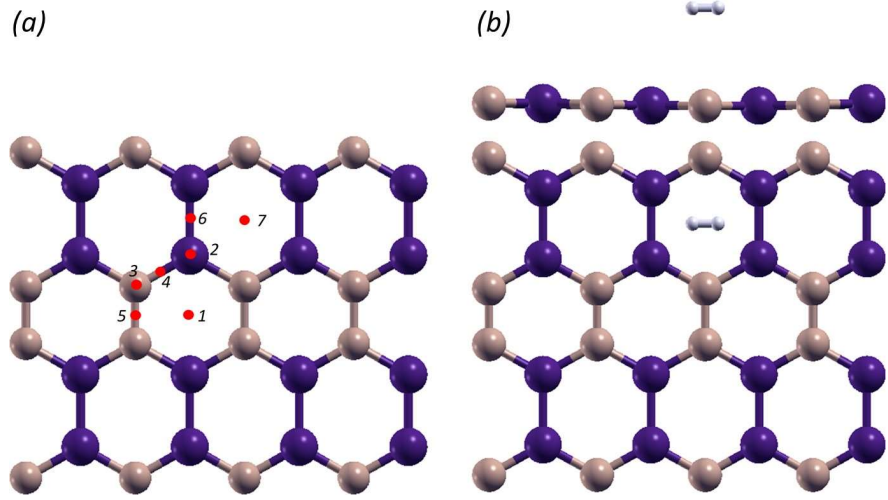

Figure 2: (a) Different initial positions of $H_2$ molecules over borophosphene, and (b) the best binding ($E_b = -0.047\ eV$) configuration.

| System | $E_b^{H_2}$ (eV/$H_2$) | $Z_{H_2-sub}$ (Å) | $R_{H-H}$ (Å) |
|---|---|---|---|
| Pristine BP | -0.047 | 2.92 | 0.751 |
| Borophene [21] | -0.045 | 2.95 | - |
| Phosphorene [34] | -0.07 | 2.98 | 0.750 |

Table 1: The binding energy ($E_b^{H2}$), height ($Z_{H2-sub}$), and bond length ($R_{H-H}$) of $H_2$ molecules for pristine 2D materials.

To elucidate the interaction between borophosphene and $H_2$ molecule, the projected density of state (PDOS) is depicted in Figure 3. Any significant interaction noticed in coherence with the low binding energy the $H_2$ molecule over borophosphene. Bader charge analysis showed that a charge of only 0.0092 e is transferred from the substrate to the $H_2$ molecule. The borophosphene substrate stability is not affected due to $H_2$ adsorption.

3.3. $H_2$ storage over decorated borophosphene

3.3.1. Decoration over borophosphene

The pristine borophosphene was first decorated by the chosen light metal adatoms (Li, Na, Ca). Motivated by the literature, adatoms were placed at two locations ($H_B$ and $H_P$) around 3 Å above the substrate as shown in Figure 4(a-c) [44, 45] and followed by the energy minimization. The binding energy of adatoms is determined by,

$$E_b^M = (E_{BP+jM} - E_{BP} - jE_M)/i. \qquad (3)$$

In the above relation, $E_{BP+jM}$ is energy of M-decorated borophosphene with $j$ adatoms, $E_{BP}$ is energy of borophosphene and $E_M$ is energy of the single adatom. The equation will result in a negative binding energy in favorable decorations. For all the adatoms, $H_B$ was found to be more stable location inferred from the stronger binding. The Li binding energies over $H_B$ and $H_P$ were computed as 1.06 eV and 0.797 eV, little higher than reported values by Du et al. [44]. For Na, those energies were 2.25 eV and 2.11 eV, and for Ca, the energies were computed as 2.77 eV, 2.51 eV, respectively. The stable configurations of decorated substrates are shown in Figure 4(a-c). The Li, Na, and Ca adatoms are located at approximately 1.64 Å, 2.07 Å, and 1.95 Å above the borophosphene substrate, respectively. The reported values in literature are summarized in Table 2.

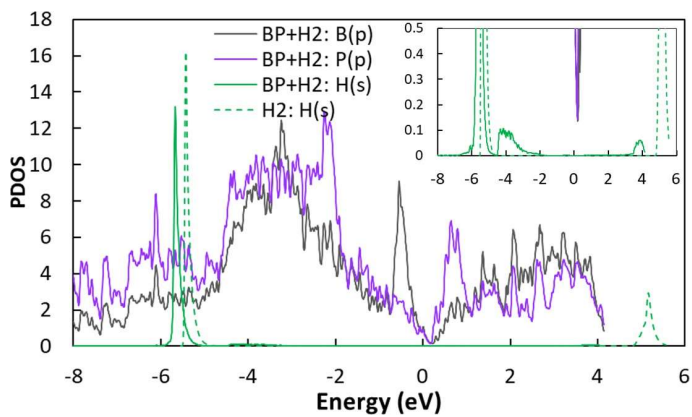

Figure 3: PDOS of B(p), P(p), and H(s) in borophosphene after hydrogen adsorption. The Fermi energy is set to zero.

Figures 4 (d-f) also show the projected density of states (PDOS) of valence electrons before and after decoration. The PDOS indicate adatom to substrate charge transfer and interaction with the substrate. When the adatoms are placed, the peaks in the conduction band are shifted toward the valence band. As observed in Figure 4 (d), the BP sheet obtained the Li(s) electron and the shift of the peak occurred closer to the Fermi energy toward the valence band due to the charge transfer. Hybridization of Li with the BP sheet is noticed at around 2.5 eV above the Fermi level. Similar affects are observed as a result of Na and Ca decoration as shown in Figures 4 (e-f) indicating ionic bonds between the substrate and the adatoms. The bader charge analysis was utilized to quantitatively determine the charge transfer. The charge transfer was calculated to be 0.88 e from Li to BP and 0.88 e from Na to BP, and 1.39 e from Ca to BP substrate. The charge transfer was reported to be 0.3 e (Hirshfeld analysis) [44] in Li decoration, 0.851 e in Na decoration [45], and 1.40 e from Ca to BP [47].

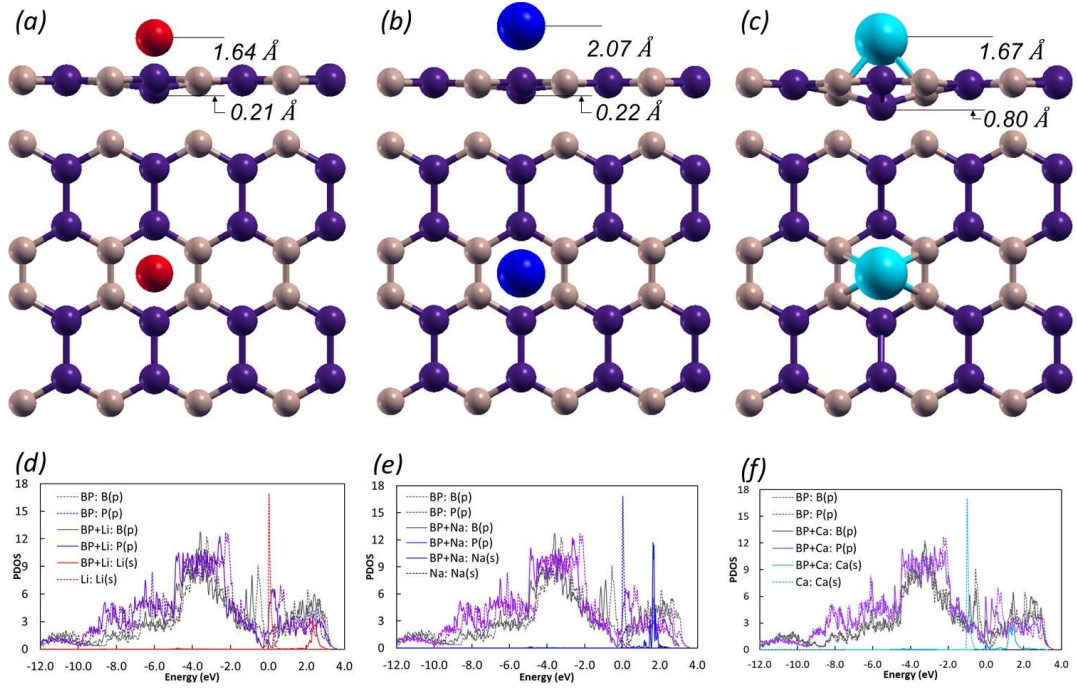

Figure 4: Stable configuration of (a) Li, (b) Na and (c) Ca decorated borophosphene and (d-f) PDOS. The Fermi energy of the respective systems are set to zero.

| Adatom | site | $E_b^M$ (eV/M) | Z (Å) | h (Å) | Δq (e) |
|---|---|---|---|---|---|
| Li | $H_B$ | 1.06 | 1.64 | 0.21 | 0.88 |
| Literature [44] | $H_B$ | 0.97 | - | - | 0.30 |
| Na | $H_B$ | 2.25 | 1.16 | 0.22 | 0.88 |
| Literature [45, 47] | $H_B$ | 0.68-0.838 | 2.04-2.19 | 0.19 | 0.851-0.88 |
| Ca | $H_B$ | 2.77 | 1.95 | 0.80 | 1.39 |
| Literature [47] | $H_B$ | 0.81 | 1.58 | 0.61 | 1.40 |

Table 2: The adatom binding energy ($E_b^M$), height from substrate (Z), corrugation height (h), and charge transfer (Δq) for different adatoms and comparison with literature.

*3.3.2. Hydrogen storage over decorated borophosphene*

To study the hydrogen storage over the decorated borophosphene substrate, $H_2$ molecules were augmented over the adatom to obtain binding energy. From Eq. 2, binding energy of the first $H_2$ on the decorated borophosphene was obtained as −0.27 eV/$H_2$, −0.20 ev/$H_2$, and −0.42 eV/$H_2$, respectively, for Li, Na, and Ca functionalization. In the relaxed system, the $H_2$ molecule was located 1.94 Å above Li with the H−H bond being 0.756 Å long. For Na and Ca decorated borophosphene, the $H_2$ was stable at a height of 2.32 Å and 2.45 Å above the adatoms with a bond length of 0.753 Å and 0.755 Å, respectively (Table 3). The stable structure of $H_2$ adsorption over decorated substrate are shown in Figure 5(a-c). The binding energies of $H_2$ over decorated borophosphene substrates are compared with borophene and phosphorene with similar decorating elements in Table 3. It can be noticed that the binding energy of $H_2$ molecule is better in borophosphene than the later ones.

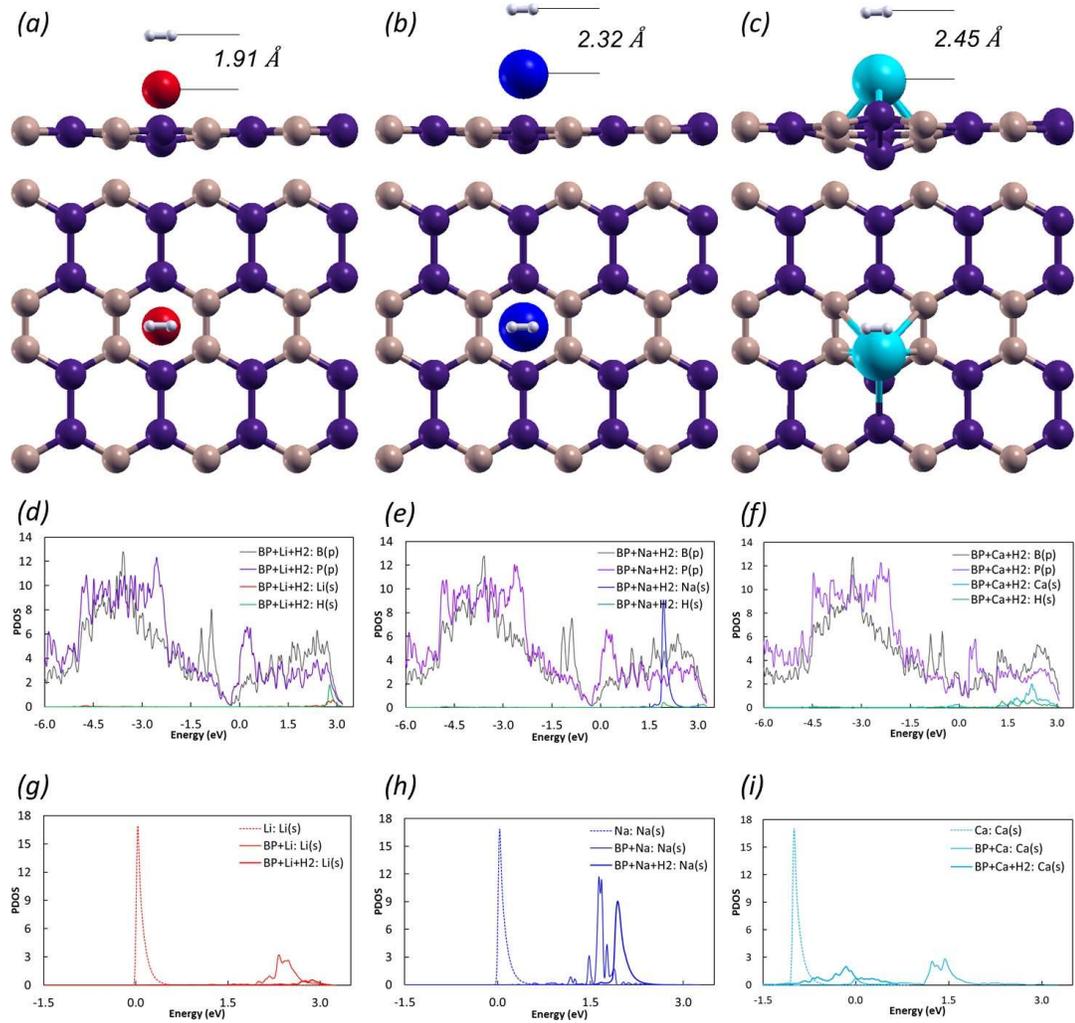

Figure 5: Stable configuration of $H_2$ adsorption over (a) Li decorated, (b) Na decorated, and (c) Ca decorated borophosphene, (d-f) relevant PDOS of the systems, (g-i) PDOS of adatoms in different systems. The Fermi energy of the respective systems are set to zero.

| System | $E_b^{H_2}$ (eV/$H_2$) | $Z_{H_2-M}$ (Å) | $R_{H-H}$ (Å) |
|---|---|---|---|
| Li-BP | -0.27 | 1.91 | 0.756 |
| Na-BP | -0.20 | 2.32 | 0.753 |
| Ca-BP | -0.42 | 2.45 | 0.755 |
| Li-Borophene [19–21] | -0.19 – -0.36 | 1.95 | 0.760 |
| Li-Phosphorene [14, 15, 18, 34] | -0.25 – -0.16 | 1.94–1.97 | 0.756–0.780 |
| Na-Phosphorene [14, 16, 58] | -0.16 – -0.158 | - | - |
| Na-Borophene [19, 21] | -0.148 – -0.34 | - | - |
| Ca-Borophene [21, 22] | -0.12 – -0.19 | 2.50 – 2.65 | 0.778 |
| Ca-Phosphorene [16, 59] | -0.12 – -0.525 | – | – |

Table 3: The binding energy ($E_b^{H_2}$), height ($Z_{H_2-M}$) from the adatom, and bond length ($R_{H-H}$) of $H_2$ molecules adsorbed over decorated borophosphene.

The PDOS of the elements after $H_2$ adsorption are presented in Figure 5 (d-f) with the Fermi energy being set to zero. The PDOS shows indicates hybridization of $H_2$ with the adatoms above Fermi energy. Due to the charge transfer from adatoms to borophosphene, the cationic adatoms (Li+, Na+, Ca+) are binding sites for the $H_2$ molecules. As a results, the $H_2$ molecules are polarized and bound by cationic adatoms through electrostatic and van der Walls interactions [29]. The PDOS of the adatoms in different systems are compared with a single isolated adatom in Figure 5 (g-i). The comparison reflects Li → substrate charge transfer during the decoration resulting in lower density of state.

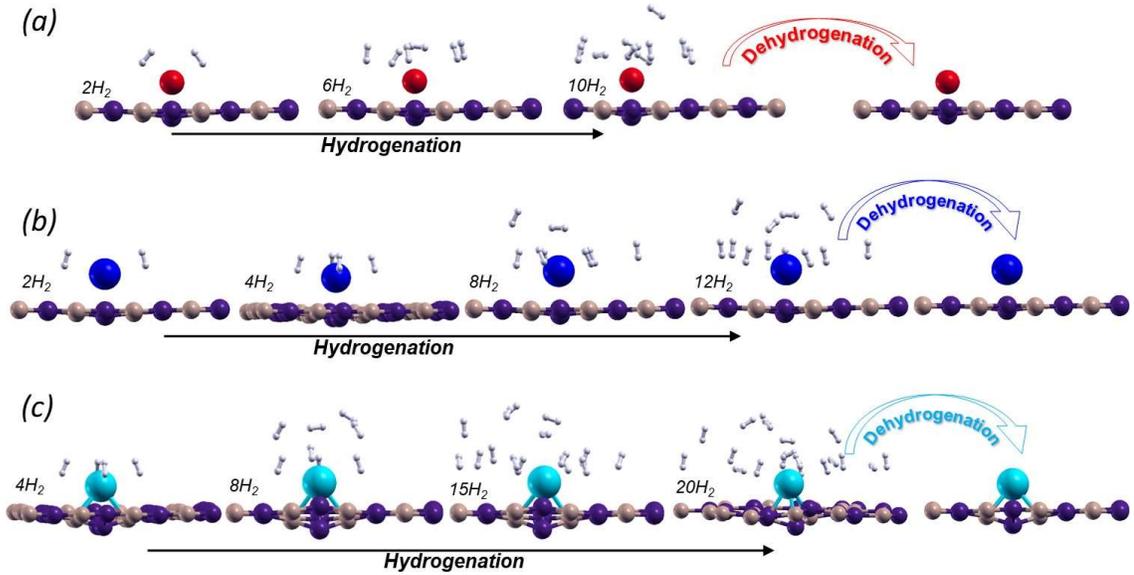

Figure 6: Hydrogenation with different numbers of molecules and dehydrogenation process of (a) Li decorated, (b) Na decorated, and (c) Ca decorated borophosphene layer.

Sequentially multiple $H_2$ molecules were added to the system to establish the highest number of $H_2$ molecules adsorbed around each adatom over the decorated borophosphene substrate. It was observed that a maximum of 10, 12, and 20 $H_2$ molecules could be adsorbed over Li, Na, and Ca decorated substrate, respectively. The hydrogenation of the decorated substrate is shown in Figure 6 for different numbers of $H_2$ molecules. The distribution of positions of $H_2$ molecules measured from adatoms is shown in Figure 7(a) along with the H–H bond lengths in Figure 7(b). The H–H bond lengths are calculated to be in the range of 0.750 – 0.765 Å. The evolutions of average per molecule binding energy during hydrogenation is shown in Figure 8(a). The bond lengths of the $H_2$ molecules and the binding energy indicate that there is no conventional Kubas type interaction [3].

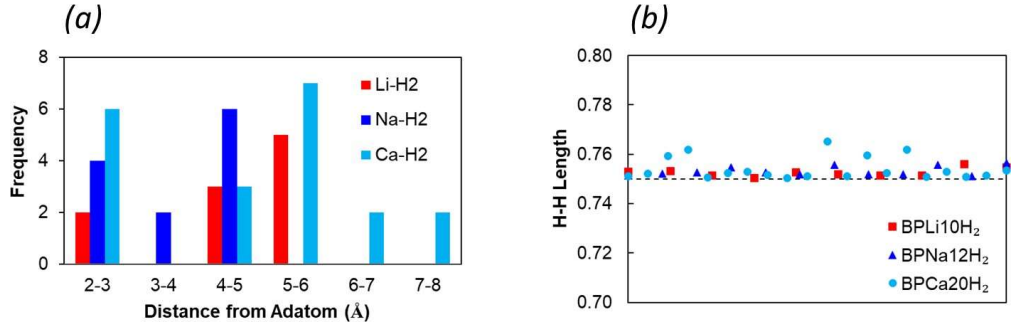

Figure 7: (a) Distribution of $H_2$ positions with respect to the adatom, and (b) H-H bond length of adsorbed $H_2$ molecules in different decorated borophosphene. The dashed line corresponds to 0.75 Å.

In addition, the desorption temperature ($T_D$) of the $H_2$ molecules is calculated from the von't Hoff equation given by [29, 60]

$$T_D = \frac{E_b}{k_b} \left[\frac{\Delta S}{R} - \ln p\right]^{-1}, \qquad (4)$$

where $E_b$ is the calculated adsorption energy (J/$H_2$). The symbols in the equation represent Boltzmann constant ($k_b$), entropy change of $H_2$ from gas to liquid ($\Delta S$ = 75.44J/mol-K), and the universal gas constant ($R$ = 8.314J/mol-K). $p$ is the equilibrium pressure taken as is 1 atm. The desorption temperature associated with the sequential adsorption is shown in Figure 8(b).

From the above calculation, the temperature for onset of desorption ($T_{DL}$) can be determined from binding energy of the last $H_2$ molecules, and by using the binding energy of the first $H_2$, the highest temperature ($T_{DH}$) for complete desorption can be obtained [60]. The temperatures for onset and full discharge was calculated as 127 – 151 K for Li decorated, 125 – 254 K for Na decorated, and 140 – 531 K for Ca decorated borophosphene, respectively (Figure 8(b)).

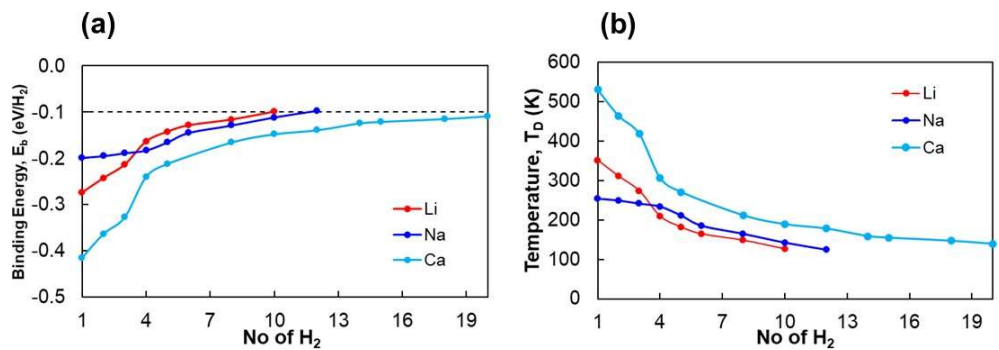

Figure 8: (a) Adsorption energy ($E_b$) and (b) Desorption temperature ($T_D$) during sequential hydrogen adsorption over functionalized borophosphene surface.

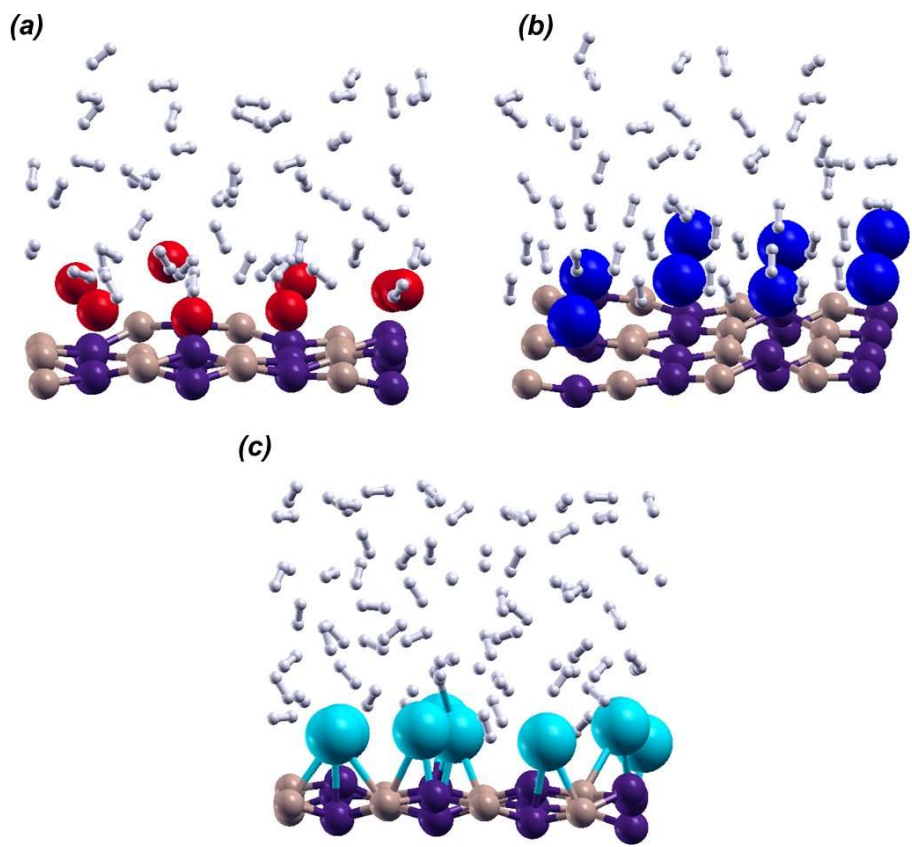

Figure 9: $H_2$ adsorption on decorated borophosphene through (a) Li overloading, (b) Na overloading, and (c) Ca overloading.

*3.3.3. Adatom overloading in borophosphene*

Overloading of the adatoms was studied through decoration at different locations with eight adatoms in the substrate. Li, Na, and Ca adatoms were adsorbed to borophosphene by an energy of 0.87 eV/Li, 1.82 eV/Na, and 2.43 eV/Ca, respectively, with the binding energies being little less than those of single adatom. Following the decoration, multiple $H_2$ molecules were than adsorbed over the decorated substrate. In the Li decorated substrate (BP+8Li), up to 48 $H_2$ molecules could be adsorbed at an average of 6$H_2$ per Li with -0.10 eV/$H_2$ as average binding energy. Within same binding energy, up to 48, and 56 $H_2$ molecules could be adsorbed over Na and Ca decorated substrates (BP+8Na, BP+8Ca), respectively (Figure 9). The gravimetric density of $H_2$ storage was computed by

$$\rho = \frac{n\, m_H}{16\, m_B + 16\, m_P + 8\, m_M + n\, m_H}, \quad (5)$$

where $m_*$ represent atomic mass of the $*$ element and $n$ being the number of $H$ atoms in the system. The gravimetric density corresponding to this adsorption is calculated to be 6.22 % for Li decorated system. However, due to larger mass of Na and Ca, resulting gravimetric density is 5.34 % and 6.08 % in Na and Ca decorated borophosphene. The gravimetric densities can be further increased through decoration on both sides of the 2D borophosphene [44, 47].

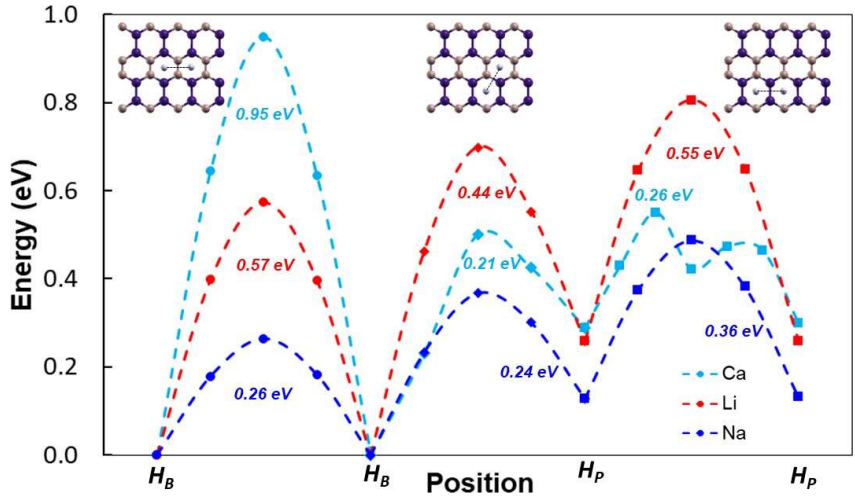

Figure 10: Diffusion barrier of adatoms across different sites along with the paths shown in inset figures.

*3.3.4. Diffusion barrier of adatoms to neighbor sites*

In adatom decoration, clustering of the adatoms with neighboring favorable is an issue. To evaluate the possibility of multiple adatoms against decoration, diffusion energy barrier from one site to the next neighbor site was calculated. Climbing image Nudged Elastic Band (CI-NEB) calculation was carried out in order to determine the energy barriers between the nearest neighbor stable locations (inset in Figure 10). The energy barriers for diffusion of Li, Na, and Ca atoms from one site to the other are depicted in Figure 10 along with the associated diffusion paths. A diffusion barrier of 0.57 eV is calculated that prohibits the Li atom to cluster from $H_B$ site to the neighboring $H_B$ site, the barrier from $H_B$ site to $H_P$ site is 0.44 eV, and that from $H_P$ site to $H_P$ site is 0.55 eV. The energy barrier values are reported in the range of 0.19 - 0.59 eV [44]). For the Na adatom, the energy barriers are obtained as 0.26 eV, 0.24 eV, and 0.36 eV for $H_B$–$H_B$, $H_B$–$H_P$, and $H_P$–$H_P$ diffusion paths (0.22 eV, 0.14 eV, and 0.31 eV reported in [45]). For Ca, the energy barriers are obtained as 0.95 eV for B to B, 0.21 eV for B to P, and 0.26 eV for those paths.

The thermal energy of the atom must be less than the energy barrier across a reaction path. The thermal energy can be computed from $E = \frac{3}{2}KT$ using the Boltzmann constant $K$ and temperature $T$. This relation results in a thermal energy of 0.071 eV at 550 K. The diffusion energy barrier of the atoms along any path is well higher than this thermal energy. It is noteworthy that the diffusion energy barriers of all the adatoms are well above the thermal energy inferring a stable decoration during $H_2$ desorption, including at the temperature associated with the full discharge ($T_{DH} = 531\ K$ for Ca).

## 4. Conclusions

Using first principles based density functional theory, hydrogen storage performance of borophosphene was investigated. Pristine borophosphene offered very weak binding (0.047 eV) that was not acceptable for efficient hydrogen storage. To enhance the storage capacity Li, Na and Ca decoration was adopted. The adatoms were with the borophosphene strongly with binding energies in the range of 1.06 – 2.77 eV. Bader charge analysis revealed a charge transfer of $0.88e – 1.39e$ from the adatom to the borophosphene sheet that contributed to the substrate-adatom interaction resulting in cationic state of the adatoms.

Binding energy of $H_2$ molecules over the decorated borophosphene was then calculated from energy minimization. The results showed that the adatom decoration significantly enhanced the $H_2$ storage capacity in comparison with pristine borophosphene. The binding energies of $H_2$ molecule over the Li, Na and Ca decorated borophosphene was calculated in the range of -0.20 to -0.42 eV/$H_2$. The $H_2$ adsorption with this range of binding energy is considered suitable for reversible storage. The PDOSs and Bader charge analysis were presented unravel the role of adatoms in charge transfer resulting in improved interaction. The binding energies of the first $H_2$ molecules over borophosphene substrates decorated with different elements were compared with reported results for borophene and phosphorene. The comparison showed that borophosphene adsorbed the $H_2$ molecules stronger than the other two with same decoration. Further, complete hydrogenation process was calculated through sequential addition of $H_2$ molecules. The results yielded adsorption of 10, 12, and 20 $H_2$ molecules at a single Li, Na, and Ca adatom over the 4 × 2 borophosphene substrates, respectively, within average binding energy of -0.10 eV/$H_2$. H–H bond lengths were found to be within 0.750–0.765 Å. Dehydrogenation temperature for complete release of all the $H_2$ molecules at atmospheric pressure was calculated using von't Hoff equation. Dehydrogenation temperature was obtained in the range of 125 – 530 K for different adatom decorated substrates.

To determine the maximum capacity of hydrogen storage, adatom overloading, i.e. decoration by several adatoms on substrate was also performed. This was pursued with 8 adatoms at the favorable sites over the supercell. The average binding energy of the adatoms reduced slightly compared to that of a single adatom. The results of $H_2$ adsorption indicated that up to 48 $H_2$ molecules were stored by Li and Na decoration, and 64 $H_2$ molecules by Ca decoration, respectively, within the average binding energy of 0.1 eV/$H_2$. Thus, decoration at single side borophosphene substrate resulted in 6.22 %, 5.34 %, and 6.08 % gravimetric density with Li, Na and Ca decoration. The gravimetric density can be further increased by decoration on both sides of the substrate. Inter-site diffusion barriers were found larger by an order of magnitude than the thermal energy at desorption temperatures indicating the clustering being unfavorable. The results may motivate further strategies to be developed for improved hydrogen storage performance in borophosphene.

**Acknowledgement**

Computational resources provided by Dr Harpreet Singh (SMS) highly appreciated.